\documentclass[12pt]{article}
\usepackage{graphicx}

\newcommand{\be}{\begin{equation}}
\newcommand{\ee}{\end{equation}}

{
\begin{document}
\begin{center} {\Large \bf Harmonic Oscillator, Coherent States,
and Feynman Path Integral}\footnote[2]{Presented at the Feynman
Festival, 23-28 August 2002, University of Maryland College Park,
Maryland, U.S.A., and also at Pac Memorial Symposium on
Theorectical Physics, 28-29 June 2002, Seoul National University,
Seoul, Korea}\footnote[3]{Supported in part by grant No.
R02-2000-00040 from the Korea Science \& Engineering Foundation.}
\end{center}
\bigskip
\begin{center}
 {Dae-Yup {Song}}
\end{center}
\begin{center}
{\it Department of Physics, University of Florida, Gainesville, FL
32611, USA\\ and Department of Physics, Sunchon National
University, Suncheon 540-742, Korea\footnote[9]{Permanent address}
}
\end{center}
\bigskip
\begin{abstract}
The Feynman path integral for the generalized harmonic oscillator
is reviewed, and it is shown that the path integral can be used to
find a complete set of wave functions for the oscillator. Harmonic
oscillators with different (time-dependent) parameters can be
related through unitary transformations. The existence of
generalized coherent states for a simple harmonic oscillator can
then be interpreted as the result of a (formal) {\em invariance}
under a unitary transformation which relates the same harmonic
oscillator. In the path integral formalism, the invariance is
reflected in that the kernels do not depend on the choice of
classical solutions.
\end{abstract}

\newpage
\section{Introduction}
Quantum mechanics can be seen as a one-dimensional field theory.
As in classical field theory, such as electrodynamics, the Green
function method is a convenient tool for the field theory. It is
therefore very tempting to try to find Green functions for quantum
mechanical systems. One of the traditional approaches to the Green
function may be in finding the complete set of modes for the
Hamiltonian (or the Schr\"{o}dinger operator) of the system.
Feynman provides a formal solution to the Green function problem:
He showed that the path integral gives the kernel and Green
function \cite{Feynman,FH}. Among the examples where path
integrals have been carried out exactly, the generalized harmonic
oscillator (GHO) is a special one, as Feynman and Hibbs have
pointed out \cite{FH}. In this contribution, the quantum mechanics
of the GHO will be discussed based on the author's works in recent
years \cite{Song,SongUni,SongPRL}. It will be shown that the
Feynman path integral carried out in terms of classical solutions
can be used to give a complete set of wave functions for the
oscillator.

In the GHO system, there exist generalized coherent (coherent and
squeezed) states which are described by classical
solutions:\footnote{This is the definition  for the generalized
coherent state adopted in this contribution. Of course, there are
many other definitions which must be connected in some way
\cite{Klauder}.} The center of probability distribution of a
coherent state moves along a classical solution, and the motion of
the width of the distribution for a squeezed state is associated
with the classical solutions, as was first found in a simple
harmonic oscillator (SHO) system \cite{Klauder,Nieto}. In recent
years, it has been shown that the harmonic oscillators of
different (time-dependent) parameters can be related through a
unitary transformation \cite{SongUni,Li}. The existence of the
generalized coherent states for a SHO can then be interpreted as a
result of the (formal) {\em invariance} under a unitary
transformation which relates the same SHO. The path integral
formalism provides a very intuitive reason for the fact that the
quantum mechanics of a GHO is described by the classical solutions
of the system, and the invariance is reflected in that the path
integral does not depend on the choice of classical solutions.

In the next section, the Feynman path integral for a quantum
mechanical system will be briefly reviewed. In Sec.~III, the path
integral will be applied to a GHO. In Sec.~IV, the unitary
relations between the different GHOs of different (time-dependent)
parameters will be discussed, and it will be argued that the
existence of generalized coherent states for a SHO is a result of
the invariance under a unitary transformation. We will also show
that the invariant (action variable) found by Lewis \cite{Lewis}
is, in the quantum theory, nothing but the transformed Hamiltonian
from that of a SHO system through the unitary transformation,
which shows why the invariant should an invariant. The final
section will be devoted to discussions.

\section{The Feynman Path Integral: A Brief Review}
Let's consider a system with the Hamiltonian,
$H_N(x_1,x_2,\cdots,x_N,p_1,p_2,\cdots,p_N)$, which gives the
Schr\"{o}dinger equation,
 \be O_N\Psi=0,
 \ee
 with the
Schr\"{o}dinger operator
 \be O_N(t,x_1,x_2,\cdots,x_N) =-i\hbar{\partial \over \partial t}+H_N.
 \ee
The Green function problem is to find the solution of the
equation:
 \be O_N(t_b,{x_1}_b,{x_2}_b,\cdots,{x_N}_b)
 G_N(b;a)
 =\delta(t_b-t_a)\prod_{i=1}^N\delta({x_i}_b-{x_i}_a).
 \ee
The kernel may be a convenient step toward the Green function, and
for a non-interacting or an isolated system it may be enough for
the analysis of the given system.  The kernel, which is a familiar
object in mathematics, is the solution of the equation:
 \be O_N(t_b,{x_1}_b,{x_2}_b,\cdots,{x_N}_b)
 K_N(b;a)=0,
 \ee
with the initial condition,
 \be K_N(b,a)\rightarrow
 \prod_{i=1}^N\delta({x_i}_b-{x_i}_a)
 ~~~{\rm as}~~~t_b\rightarrow t_a+0.
 \ee

Feynman provides a formal solution for the kernel: He showed that
the kernel can be obtained through the path integral
 \be K(b,a)=\int_{{x_1}_a}^{{x_1}_b}
 \cdots\int_{{x_N}_a}^{{x_N}_b}
 \exp[{i\over \hbar}\int_{t_a}^{t_b} L(t,x_1,\cdots,x_N)]{\cal D}
 x_1\cdots{\cal D} x_N.~~~~~
 \ee
In his book with Hibbs \cite{FH}, he then asserted that the Green
 function can be written as
 \begin{eqnarray}
G(b,a)=\left\{
  \begin{array}{l l}
        K(b,a) &{\rm if }~~~ t_b> t_a, \\
        0 &{\rm if }~~~ t_b< t_a .\\
  \end{array}
  \right.
\end{eqnarray}

Assuming a complete set of wave functions
$\{\psi_{n_1,n_2,\cdots,n_N}\}$ for the $O_N$, an alternative
expression for the kernel satisfying Eqs. (4,5) is given as
 \begin{eqnarray}
 &&K(b,a)\cr
 &&=\sum_{n_1,\cdots,n_N}\psi_{n_1,n_2,\cdots,n_N}(t_b,{x_1}_b,\cdots,{x_N}_b)
 \psi_{n_1,n_2\cdots,n_N}^*(t_a,{x_1}_a,\cdots,{x_N}_a).~~~~~~~
 \end{eqnarray}
If the exact kernel can be found, the expression in Eq.~(8) may be
used to find a complete set. If we do not require $K(b,a)=0$ for
$t_b<t_a$, in this expression, there is a relation:
 \be K^*(b,a)=K(a,b).
 \ee
Though the kernel has been studied for various models through the
path integral or the mode summation method, the list of the cases
where the exact kernel is known is still limited
\cite{IKG,Schulman}.

\section{A Generalized Harmonic Oscillator and the Path Integral}

As shown by Feynman and Hibbs, \cite{FH}, the path integral for a
general quadratic system can be carried out, and the kernel is
written in terms of the classical action up to a term which purely
depends on the initial time $t_a$ and the final time $t_b$ of the
path integral. A Lagrangian of an $N$-dimensional quadratic system
may be written as
 \begin{eqnarray}
 L&=&\sum_{i=1}^N\left( {1\over 2} M(t) \dot{x}_i^2
 - {1\over 2}M(t)w^2(t) x_i^2 +F_i(t) x_i \right.\cr
&&~~~~~~\left. + {d \over dt} (M(t)a(t)
 x_i^2) + {d \over dt} (b_i(t) x_i)\right) +f(t),
 \end{eqnarray}
while, for $N=1$, it gives a general quadratic system in one
dimension. This Lagrangian clearly shows that the classical
equation of motion of the quadratic system is same to that of a
GHO of time-dependent mass and frequency with an external force:
\be
 {d \over {dt}} (M \dot{{x}_i}) + M(t) w^2(t) {x}_i =F_i(t),
 \ee
since the terms written as a total derivative with respect to time
do not affect the classical equation of motion. Indeed, the
quantum mechanics of the GHO has been a subject of intensive study
over a long period, mainly relying upon the invariant (See Sec.
IV).

In addition to the observation by Feynman and Hibbs, the fact that
the kernel should satisfy the Schr\"{o}dinger equation and the
initial condition can be used to determine the kernel uniquely for
the GHO \cite{Song}. If the two linearly independent homogeneous
solutions of Eq.~(11) are denoted as $u(t),~v(t)$ with a
particular solution ${x_i}_p(t)$, then one can find that the
kernel is written as
\begin{eqnarray}
 &&K(b,a) \cr
 &&=\left({1\over 2i\pi\hbar}{\Omega \over v(t_b)u(t_a)-u(t_b)v(t_a)}
     \right)^{N/2}\exp[{i\over \hbar}\int_{t_a}^{t_b}f(z)dz]\cr
 &&\times\exp\left[{i\over \hbar}\left\{M(t_b)a(t_b)\vec{r}_b^2
    +\vec{b}(t_b)\cdot\vec{r}_b
      +M(t_b)\vec{r}_p(t_b)\cdot{\vec{r}}_b +\xi(t_b)-(a\leftrightarrow b)\right\}
      \right]\cr
 &&\times\exp[{i\over 2\hbar(v(t_b)u(t_a)-u(t_b)v(t_a))}   \cr
 &&~~~~~~~\times
 \left\{M(t_a)(u(t_b)\dot{v}(t_a)-\dot{u}(t_a)v(t_b))(\vec{r}_a-\vec{r}_p(t_a))^2 \right.\cr
 &&~~~~~~~~~~~~-\Omega(\vec{r}_a-\vec{r}_p(t_a))\cdot(\vec{r}_b-\vec{r}_p(t_b))
   \left.\left.+(a\leftrightarrow b)\right\}\right],
 \end{eqnarray}
where $\vec{r}=(x_1,x_2,\cdots,x_N)$,
$\vec{r}_p=({x_1}_p,{x_2}_p,\cdots,{x_N}_p)$. In Eq.~(12),
$\xi(t)$ and a time-constant $\Omega$ are defined as
 \begin{eqnarray}
 \dot{\xi}(t)&=&\sum_{i=1}^N {1\over
 2}(Mw^2{x_i}_p^2-M\dot{x_i}_p^2),\\
 \Omega &=&M(t)(u(t)\dot{v}(t) -v(t)\dot{u}(t)).
 \end{eqnarray}
For the case ${x_i}_p(t_a)=0$, if we replace
$u(t_a)v(t)-v(t_a)u(t)$ with $v_s(t)$, the kernel $K$ in Eq.~(12)
can be straightforwardly obtained from the kernel of the
one-dimensional harmonic oscillator \cite{Song}. Indeed, it has
been known that the kernel does not depend on the choice of
classical solutions \cite{Song}.

The form of the kernel suggests that the wave functions can also
be written in terms of the classical solution. Explicitly, making
use of the expression in Eq.~(8), a complete set of wave functions
$\{\psi_{n_1,n_2,\cdots,n_N}|$ $n_1,n_2,\cdots,n_N
=0,1,2,\cdots\}$ can be read out from the kernel:
 \begin{eqnarray}
 &&\psi_{n_1,n_2,\cdots,n_N}(t,{x_1},x_2\cdots,{x_N})\cr
 &&=({\Omega \over \pi\hbar\rho^2(t)})^{N/ 4}
    \exp{[{(\vec{r}-\vec{r}_p(t))^2\over 2\hbar}(-{\Omega \over\rho^2(t)}
               +i M(t){\dot{\rho}(t) \over \rho(t)})]}  \cr
 &&~\times
 \exp[{i\over\hbar}(\xi(t)+M(t)a(t)\vec{r}^2+
     (M(t)\dot{\vec{r}}_p(t)+\vec{b}(t))\cdot\vec{r}+\int^t f(z)dz)]
 \cr
 &&~\times\prod_{i=1}^N {1\over \sqrt{2^{n_i} {n_i}!}}
     [{u(t)-iv(t) \over \rho(t)}]^{n_i+{1\over 2}}
         H_{n_i}(\sqrt{\Omega \over \hbar} {x_i -{x_i}p(t) \over
         \rho(t)}),
\end{eqnarray}
with
 \be \rho(t)=\sqrt{u^2(t)+v^2(t)},\ee
 as in the one-dimensional case \cite{Song,SongPRL,Ji}.

 The Hamiltonian corresponding to the Lagrangian in Eq.~(10) is written,
 in terms of the operators $\vec{r}$ and
$\vec{p}$,  as
 \begin{eqnarray}
 H &=& {\vec{p}^2 \over 2 M(t)} - a(t)[\vec{p}\cdot\vec{r}+\vec{r}\cdot\vec{p}]
     +{1\over 2} M(t)c(t)\vec{r}^2 \cr
   &&-{\vec{b}(t)\over M(t)}\cdot\vec{p}+\vec{d}(t)\cdot\vec{r}
      +( {\vec{b}^2(t) \over 2M(t)} -f(t)),
 \end{eqnarray}
where
 \be
 c(t)=w^2 + 4a^2 -2 \dot{a} -2 {\dot{M}\over M}a, ~~~
 \vec{d}(t)= 2a\vec{b} -\dot{\vec{b}}(t) -\vec{F}(t).
 \ee
Though the terms proportional to $a(t)$ and $\vec{b}(t)$ make the
Hamiltonian look rather complicated, from the wave function in
(15), it can be found that such terms can be removed from the
system through a simple unitary transformation \cite{Song}. For
the one-dimensional case, the quantum description of a general
quadratic system can therefore be found, through a unitary
transformation, from that of a GHO system. From now on we only
consider the case of $N=1$ \cite{Comment}. We also take
$a(t),{b}(t),f(t)$ to vanish, since the general case can be
recovered through a simple unitary transformation.

\section{Generalized Coherent States from the Unitary Invariance}
An historical account of the generalized coherent states of the
simple harmonic oscillator (SHO) system has been well summarized
by Klauder and Skagerstam \cite{Klauder} and by Nieto
\cite{Nieto}. In recent years, it has been shown that harmonic
oscillators of different (time-dependent) parameters can be
related through a unitary transformation \cite{Li,SongUni}. In
this section, it will be shown that the existence of generalized
coherent states in a SHO system can then be interpreted as a
result of the (formal) {\em invariance} under a unitary
transformation which relates the same SHO \cite{SongUni}.

The generalized harmonic oscillator (GHO) of time-dependent mass
and frequency with an external force may be described by the
Hamiltonian:
 \be
 H_F ={p^2\over 2M(t)}+{1\over 2}M(t)w^2(t)x^2-F(t) x=H-F(t) x.
 \ee
The Hamiltonian for the SHO system of unit mass and unit frequency
may be given by
 \be H_s={p^2 \over 2}+{x^2 \over 2}. \ee
We define the three different Shr\"{o}dinger operators
 \be
 O_F(t)=-i\hbar{\partial\over \partial t} +H_F,~~~
 O(t)=-i\hbar{\partial\over \partial t} +H,~~~
 O_s(\tau)=-i\hbar{\partial\over \partial \tau} +H_s,
 \ee
which correspond to the Hamiltonian $H_F,~ H,$ and $H_s$,
respectively.

With $x_p$ being a particular solution of the classical equation
of motion for the system described by $H_F$,  as in the previous
section, we define $\xi$ through the relation
 $\dot{\xi}(t)={1\over 2}(Mw^2{x}_p^2-M\dot{x}_p^2).$
A unitary operator defined by
 \be U_F= e^{{i \over\hbar}\xi} \exp[{i\over \hbar}M\dot{x}_px]
 \exp[-{i\over \hbar}x_p p]
 \ee
then gives the relation
 \be U_F O(t) U_F^\dagger =O_F(t)
 \ee
connecting the systems  described by $H$ and $H_F$. If $\tau$, the
time of the SHO system, and t, the time of the system described by
$H$, are related by
 \be d\tau={\Omega \over M(t)\rho^2(t)} dt,
 \ee
then the unitary operator given by
 \be U_S=\exp[{i \over 2 \hbar}M(t)
 {\dot{\rho}(t) \over \rho(t)}x^2] \exp[-{i\over 4\hbar}\ln({\rho^2(t) \over
 \Omega})(xp+px)]
 \ee
also connects the system described by $H_s$ and $H$, as
 \be U_S O_s(\tau)U_S^\dagger\mid_{\tau=\tau(t)}= {M \rho^2
 \over \Omega} O=\left( {dt \over d\tau}\right)O.
 \ee

The two unitary relations given in Eqs.~(23,26) and the unitary
transformation mentioned in the previous section thus relate the
SHO system of unit mass and unit frequency to a general quadratic
system, provided the solutions of classical equation of motion of
the quadratic systems are known. For simplicity, we only consider
the relations between the systems described by $H_F$ and $H_s$ in
detail.

If $\phi_s(x)$ is an eigenstate of Hamiltonian $H_s$ satisfying
$H_s=E\phi_s(x)$, then from the unitary relations, a wave function
for the system $H_F$, satisfying $O_F(t)\psi=0$, is given by
\begin{eqnarray}
\psi(t,x)&=&e^{-iE\tau/\hbar}\mid_{\tau=\tau(t)}
         U_FU_S\phi_s(x)\\
&=&({\Omega \over \rho^2})^{1/4}
     ( {u(t)-iv(t) \over \rho(t)} )^{E/\hbar}e^{{i \over\hbar}\xi}
     \exp[{i\over \hbar}M\dot{x}_px] \cr
 &&\times
     \exp[{i \over 2 \hbar}M {\dot{\rho} \over \rho} (x-x_p)^2] )
     \phi_s(\sqrt{\Omega \over \rho^2}(x-x_p)).
\end{eqnarray}
 From the well-known eigenfunctions of the SHO system, then a complete set for the
 system described by
$H_F$ is therefore given as $\{\psi_n (t,x)|n=0,1,2\cdots\}$,
where
 \begin{eqnarray}
 \psi _n (t,x) &=& \frac{1}{{\sqrt {2^n n!} }}
  (\frac{\Omega }{{\pi \hbar}})^{1/4} \frac{1}{{\sqrt \rho  }}
  \left(\frac{{u - iv}}{\rho }\right)^{n +1/2}
  {\rm{exp[}}\frac{{\rm{i}}}{\hbar }(M\dot{x}_p x + \xi  )]\cr
 && \times {\rm{exp[}}\frac{{{\rm{(}}x - x_p )^2 }}{{2\hbar }}( -
 \frac{\Omega }{{\rho ^2 }} + iM\frac{{\dot \rho }}{\rho })]
 {\rm{H}}_n (\sqrt {\frac{\Omega }{\hbar }} \frac{{x - x_p
 }}{\rho }).
 \end{eqnarray}

For the case where $M(t)=1=w(t)$ and $F(t)=0$, the unitary
relations in Eqs.~(23,26) become the desired relations between the
same system. As one example, if we choose the classical solutions
as $u(t)=\cos t,~v(t)=\sin t$ and $x_p(t)=0$, then $U_S=U_F=1$.
However, there are choices  which give non-trivial $U_S$ and/or
$U_F$; for example, the choice $u(t)=\cos t,~v(t)=\sin t$, and
$x_p(t)=\cos t$ gives $U_S=1$ with a non-trivial $U_F$, so that
$\psi _n (t,x)$ describes a coherent state whose probability
distribution moves along with $x_p(t)~~~(=\cos t)$.\footnote{Note
that, even for $F(t)=0$, a non-trivial $x_p$ can be chosen
\cite{SongUni}.} On the other hand, if we choose $u(t)=\cos
t,~v(t)=C\sin t$ $~(C\neq 1)$, with $x_p(t)=0$, then $U_F=1$ but
$U_S$ gives a squeezed state whose probability distribution
pulsates periodically with the period 1. In this way, it is clear
that the presence of generalized coherent states in SHO systems is
a result of the invariance of the Schr\"{o}dinger operator under
the unitary transformation [Due to the relation in Eq.~(26), this
invariance is up to a time-scaling given in Eq.~(24) and a
multiplication of a purely time-dependent term].

An operator $I$ which has long been of interest for the GHO can be
obtained from $H_s$ as
 \be I=U_FU_S H_sU_S^\dagger U_F^\dagger.
 \ee
By comparing the Schr\"{o}dinger operators, one can find that
 \be
H_F=\left({d\tau \over dt}\right)I+U_FU_S (-i\hbar{\partial \over
\partial t}) U_S^\dagger U_F^\dagger,
 \ee
The expression of $H_F$ given in Eq.~(31) can be used to show that
$I$ is an invariant of the system of $H_F$ satisfying
 \be i\hbar{\partial \over \partial t} I=[H_F,I].
 \ee
The explicit expression of $I$ is given as \cite{SongPRL}
\begin{eqnarray}
I&=&{1 \over 2 \Omega}
   [{\Omega^2 \over \rho^2}(x-x_p)^2  \cr
 & & + \{M\dot{\rho}(x-x_p)- \rho(p-M\dot{x}_p)\}^2].
\end{eqnarray}
If we simply take the operators $x$ and $p$ as the canonical
variables of classical physics, one can find that $I$ is the
action variable satisfying
 \be
 {dI\over dt}= {\partial I\over \partial t} +[I, H]_{PB}=0.
 \ee

For $x_p=0$, the expression of $I$ has first been found by Lewis
\cite{Lewis} through the asymptotic theory of Kruskal. As far as
the quantum theory of GHO is concerned, the unitary transformation
method gives a simple way to find the invariant and clearly shows
the reason why the invariant is in fact an invariant.

\section{Discussions}
The Feynman path integral for the GHO has been reviewed, to show
that the path integral can be used to find a complete set of wave
functions for the oscillator. Harmonic oscillators with different
parameters have been shown to be related through unitary
transformations, and the existence of generalized coherent states
for a SHO has been interpreted as a result of the (formal) {\em
invariance} under a unitary transformation which relates the same
SHO. It has been shown that the unitary transformation applied to
the Hamiltonian of a SHO gives an invariant of the time-dependent
system; this feature will also be true for any time-dependent
system which is related to a time-independent system through a
unitary transformation. This invariant becomes an exact action
variable if the operators are considered as classical variables,
while the harmonic oscillator is also interesting in that it
provides a fruitful ground to study the relations between
classical and quantum physics \cite{KW,SongPRL}.

For the harmonic oscillator with an additional inverse-square
potential, it has been known that the path integral can be carried
out exactly \cite{KL}. For this case, it has been shown that a
unitary transformation method could be used to find a complete set
from a simple case \cite{SongUni}. The unitary method can also be
applied for the $N$-body harmonic oscillators interacting through
inverse-square potential (Calogero-Sutherland Model)
\cite{SongMany}, while the kernel has not been known for this
case. Indeed, it would be very interesting, if we could evaluate
the kernel for an interacting many-body system.

There could be different representations for the same model, as
the representation of a GHO system is determined by the choice of
classical solutions. However, I believe that the kernel will be
unique for a given model. A simple anyon model may be realized as
a non-interacting system of two-body harmonic oscillators in two
dimension obeying fractional statistics. Though this model with
fractional statistics is described by the same Hamiltonian of the
non-interacting harmonic oscillators with usual statistics, the
kernels for the two systems are different, which shows that the
statistics determine the model in this case \cite{LM,Wilczek}.

Last, but not least, we have to add that the solutions presented
in this contribution are still formal, in that the explicit wave
functions and kernel will be given once the classical solutions
are found. Some general features of the classical solutions have
been discussed in Ref.~\cite{SongBerry}. In general, the classical
problems are not trivial at all and amount to solving the
one-dimensional, time-independent Schr\"{o}dinger equation with an
arbitrary potential.

\section*{Acknowledgments}
The author thanks Professor John R. Klauder for a careful reading.

\end{document}